\documentclass[aps,prl, twocolumn,showpacs,preprintnumbers]{revtex4}

\usepackage{graphicx}
\usepackage{dcolumn}
\usepackage{bm}
\usepackage{amsmath, amssymb}
\usepackage{latexsym}

\def\fund{  \> {\vcenter  {\vbox
               {\hrule height.6pt
                \hbox {\vrule width.6pt  height5pt
                      \kern5pt
                      \vrule width.6pt  height5pt}
                \hrule height.6pt}
                         }
               }
            \>\>  }

\def\afund{  \> \overline{ {\vcenter  {\vbox
               {\hrule height.6pt
                \hbox {\vrule width.6pt  height5pt
                      \kern5pt
                      \vrule width.6pt  height5pt}
                \hrule height.6pt}
                         }
               } }
            \>\>  }

\def\sym{  \> {\vcenter  {\vbox
              {\hrule height.6pt
               \hbox {\vrule width.6pt  height5pt
                      \kern5pt
                      \vrule width.6pt  height5pt
                      \kern5pt
                      \vrule width.6pt  height5pt}
               \hrule height.6pt}
                         }
               }
            \>\>  }

\def\symbar{  \> \overline{ {\vcenter  {\vbox
              {\hrule height.6pt
               \hbox {\vrule width.6pt  height5pt
                      \kern5pt
                      \vrule width.6pt  height5pt
                      \kern5pt
                      \vrule width.6pt  height5pt}
               \hrule height.6pt}
                         }
               } }
            \>\>  }

\def\asym{ \> {\vcenter  {\vbox
                 {\hrule height.6pt
                  \hbox {\vrule width.6pt  height5pt
                         \kern5pt
                         \vrule width.6pt  height5pt}
                  \hrule height.6pt
                  \hbox {\vrule width.6pt  height5pt
                         \kern5pt
                         \vrule width.6pt  height5pt}
               \hrule height.6pt}
                         }
               }
            \>\>  }

\def\asymbar{ \> \overline{ {\vcenter  {\vbox
                 {\hrule height.6pt
                  \hbox {\vrule width.6pt  height5pt
                         \kern5pt
                         \vrule width.6pt  height5pt}
                  \hrule height.6pt
                  \hbox {\vrule width.6pt  height5pt
                         \kern5pt
                         \vrule width.6pt  height5pt}
               \hrule height.6pt}
                         }
               } }
            \>\>  }

\begin{document}

\pagestyle{plain}

\preprint{RIKEN-TH-31}

\title{Color Superconductivity from Supersymmetry}

\author{Nobuhito Maru}
 \email{maru@riken.jp}
\author{Motoi Tachibana}%
 \email{motoi@riken.jp}

\affiliation{%
Theoretical Physics Laboratory, RIKEN, 
Wako, Saitama, 351-0198, Japan
}%

\date{\today}
\begin{abstract}
A supersymmetric composite model of color superconductivity is proposed. 
Quarks and diquarks are dynamically generated as composite fields
by a newly introduced strong gauge dynamics. 
It is shown that the condensation of 
the scalar component of the diquark supermultiplet occurs 
when the chemical potential becomes larger than some critical value. 
We believe that the model well captures 
aspects of the diquark condensate behavior 
and helps our understanding of the diquark dynamics in real QCD. 
The results obtained here might be useful when we consider a theory 
composed of quarks and diquarks. 
\end{abstract}
\maketitle

\begin{figure}
\begin{center}
\includegraphics[width=7cm,height=4cm]{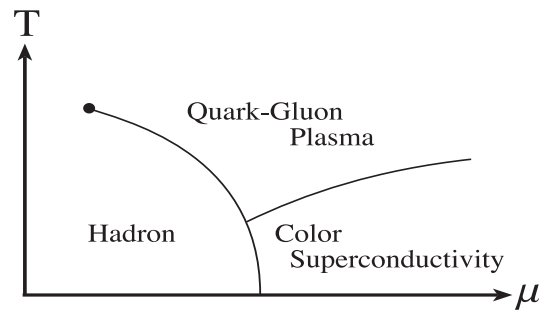}
\end{center}
\caption{\label{fig:phase} 
(Conjectured) phase diagram of hot and dense quark matter.}
\end{figure}

In the past few years, interesting properties of quark matter
with nonzero temperature and baryon density such as the phase diagram 
in Fig.~\ref{fig:phase} have been extensively studied. 
In particular, color superconductivity \cite{CSC}
is one of the most promising possiblities to occur 
in such a system at low temperature because there exists an 
attractive interaction between quarks on the Fermi surface through
the antisymmetric color $\bar{3}$ gluon exchange and it results in
the quark-quark (not quark-antiquark) pairing, 
so-called {\it diquark condensate}.
So far, a lot of people have tried to investigate this 
phenomena using different methods. At asymptotically 
high density, one can perform the weak coupling analysis of QCD 
and compute the superconducting gap \cite{son}. On the other hand, 
at lower densities where the quark chemical potential $\mu$ is
of order 500 MeV, since QCD itself is not tractable in this strong
coupling regime and in addition there is a nortorious 
sign problem on the lattice Monte Carlo simulation with
nonzero $\mu$, only some model studies (for instance, 
the Nambu-Jona-Lasinio model) which mimic crucial features of QCD
such as chiral symmetry have been done so far \cite{NJL}.
According to these model studies, a phase transition between
hadronic phase and color superconducting one is suggested
to happen. Combining this 
with the result of the weak coupling analysis,
we expect a similar phase structure is realized in dense QCD. 

In the light of these current situations, the purpose of this
paper is to propose some new way of understanding a system of
dense quark matter based on {\it supersymmetric} (SUSY) QCD.
In SUSY gauge theories, 
some exact nonperturbative results,
which are powerful tools to study the strong coupling dynamics,
have been already known \cite{review}. 
It might be therefore interesting to investigate 
color superconductivity using SUSY gauge dynamics and consider 
whether we can obtain some insight into real QCD.

In \cite{HLM}, symmetry breaking pattern was studied
by using the exact results of SUSY QCD with nonzero chemical potetnial
and was compared to that obtained from the analysis via
nonsupersymmetric QCD \cite{schafer}. 
One of the important observations is 
that the chemical potential can be incorpolated as 
the time component of a fictitious gauge field of 
the $U(1)_B$ baryon number symmetry at zero temperature, 
which leads to a tachyonic SUSY breaking scalar mass. 
However, gauge {\em variant} quantities such as the diquark 
degrees of freedom was not treated in their model. 
In order to see what happens to the system at large chemical
potential, however, as has been already mentioned before, 
it is indispensable to include that degrees of freedom.
Therefore in this paper we try to extend the work of \cite{HLM}
so as to involve gauge variant quantities at an intermediate
energy range. 

Along these line of thought, 
we propose 
{\em a supersymmetric composite model of color superconductivity}, 
in which quarks and diquarks appear as massless composites 
at low energy by a newly introduced strong coupling gauge dynamics, 
not by QCD dynamics. 
We find a certain parameter region where the scalar component of 
diquark supermultiplet condensation occurs 
when the chemical potetnial gets larger than some critical value. 
Although our model is not fully realistic 
in the sense that not diquarks themselves 
but the scalar component of the diquarks supermultiplet condense, 
nevertheless we believe that our model well captures some important 
aspects of the diquark condensate and  
helps our understanding for the color superconductivity in real QCD. 
The results obtained here may give an interesting insight on the
phase structure at the intermediate region of the quark chemical
potential.


Let us explain our model, 
which is based on a ${\cal N}=1$ SUSY $SO(N= N_f + 4)$ gauge theory 
with $N_f$ vector representations \cite{IS}. 
Its non-Abelian global symmetry $SU(N_f)$ is extended to 
$SU(3)_C \times SU(N_f)_L \times SU(N_f)_R$ where 
$SU(3)_C$(usual color symmetry) gauge theory is assumed to be weakly gauged 
compared with $SO(N)$ gauge theory, 
${\i.e.}\Lambda_{SO(N)} > \Lambda_{SU(3)_C}$ for the dynamical scales 
of each gauge group. 
Matter content is summarized below. 
\begin{eqnarray}
\label{elementary}
Q &=& (\fund, \fund, \fund, {\bf 1})_{1,1,1}, \\
\bar{Q} &=& (\fund, \afund, {\bf 1}, \afund)_{-1,1,1}, \\
X &=& (\fund, {\bf 1}, {\bf 1}, {\bf 1})_{0, -6 N_f, 3 - N}
\end{eqnarray}
where the representations in the parenthesis are transformation properties 
under the group $SO(N) \times SU(3)_C \times SU(N_f)_L \times SU(N_f)_R$, 
where $N = 6 N_f + 5$ in the present case. 
The numbers in the subscripts are charges for nonanomalous $U(1)$ 
global symmetries $U(1)_B \times U(1)_A \times U(1)_R$, 
which each $U(1)$ symmetries are linear combinations of 
the original anomalous $U(1)$ symmetries. 
$U(1)_B$ is a baryon number symmetry which plays an impotant role 
for considering the chemical potential effects. 
$U(1)_A$ is a non-R symmetry and $U(1)_R$ is an R-symmetry.

$SO(N)$ gauge theory under consideration is asymptotically free 
and known to be in a confining phase at the infrared (IR) \cite{IS}. 
At the scale $\Lambda_{SO(N)}$, 
the theory becomes strongly coupled and $SO(N)$ gauge invariant 
composite fields appear 
as massless degrees of freedom in the low energy effective theory. 
\begin{eqnarray}
\label{comp1}
(Q^2) &=& (\afund, \asym, {\bf 1})_{2, 2, 2}, (\sym, \sym, {\bf 1})_{2,2,2}, \\
(\bar{Q}^2) &=& (\fund, {\bf 1}, \overline{\asym})_{-2, 2, 2}, 
(\overline{\sym}, {\bf 1}, \overline{\sym})_{-2,2,2}, \\
(X^2) &=& ({\bf 1}, {\bf 1}, {\bf 1})_{0, -12N_f, 6 - 2N}, \\
(QX) &=& (\fund, \fund, {\bf 1})_{1, 1 - 6N_f, 4 - N}, \\
(\bar{Q}X) &=& (\afund, {\bf 1}, \afund)_{-1, 1 - 6N_f, 4 - N}, \\
\label{comp2}
(Q\bar{Q}) &=& ({\bf 1}, \fund, \afund)_{0,2,2}, 
({\bf 8}, \fund, \afund)_{0, 2, 2}
\end{eqnarray}
where the representations in the parenthesis are 
those under the group $SU(3)_C \times SU(N_f)_L \times SU(N_f)_R$. 
The numbers in the subscripts are charges for nonanomalous $U(1)$ symmetries 
$U(1)_B \times U(1)_A \times U(1)_R$. 
Note that $Q^2(\bar{Q}^2)$ has symmetric (its conjugate) and anti-symmetric 
(its conjugate) representations 
under $SU(3)_C \times SU(N_f)_L \times SU(N_f)_R$ 
because $SO(N)$ indices are contracted 
symmetrically and the superfields are bosonic. 
The anti-symmetric ones correspond to ``diquark" superfield 
responsible for the condensation 
when the chemical potential becomes larger than some critical value. 
$QX(\bar{Q}X)$ correspond to usual quarks (anti-quarks) superfields. 
Thus, quarks and diquarks coexist as massless composites in the low energy. 
As a nontrivial check that composite fields (\ref{comp1})--(\ref{comp2}) 
are appropriate massless degrees of freedom, 
we can easily show that 
the 't Hooft anomaly matching conditions for $[SU(N_f)_{L,R}]^3$, 
$[SU(N_f)_{L,R}]^2 U(1)_{B,A,R}$, $U(1)_{B,A,R}$, $[U(1)_{B,A,R}]^3$, 
$[U(1)_{B,A}]^2 U(1)_R$, 
$[U(1)_{A,R}]^2 U(1)_B$, $[U(1)_{B, R}]^2 U(1)_A$ 
at the origin of the moduli space are satisfied 
between elementary fields $Q, \bar{Q}, X$ and 
all composite fields (\ref{comp1})--(\ref{comp2}). 

The low energy effective superpotential is generated 
by the gaugino condensation in the unbroken gauge group 
$SO(N) \to SO(4) \simeq SU(2)_L \times SU(2)_R$;
\begin{eqnarray}
\label{superpot}
W_{{\rm eff}} 
= 2(\epsilon_L + \epsilon_R)
\left( \frac{\Lambda_{SO(N)}^{6N_f+4}}{[{\rm det}(Q\bar{Q})] X} 
\right)
\end{eqnarray}
where $\epsilon_{L,R} = \pm 1$ are phase factors reflecting 
the number of SUSY vacua suggested from Witten index \cite{Witten}.

Since our interest is whether the condensation of the scalar component 
of the diquark supermultiplet occurs or not
as the chemical potential changes, we need to estimate 
the soft scalar mass squareds for composite fields, 
whose sign indicate whether composite fields develop 
vacuum expectation values (VEVs) or not. 
In softly broken SUSY gauge theory, 
it is well known that soft scalar masses in the IR region can be derived 
from those in the ultraviolet (UV) region by the procedure in Ref.~\cite{AR}. 
The effective K\"ahler potential for composite fields is fixed by symmetries 
and the renormalization group (RG) invariance, 
\begin{eqnarray}
\label{effkahler}
K_{{\rm eff}} &=& c_{(Q^2)} \frac{{\cal Z}_{(Q^2)}}{I} (Q^2)^\dag e^{2V_B} (Q^2) 
\nonumber \\
&+& c_{(\bar{Q}^2)} \frac{{\cal Z}_{(\bar{Q}^2)}}{I} 
(\bar{Q}^2)^\dag e^{-2V_B} (\bar{Q}^2) 
\nonumber \\
&+& c_{(X^2)} \frac{{\cal Z}_{(X^2)}}{I} (X^2)^\dag (X^2) \nonumber \\
&+& c_{(QX)} \frac{{\cal Z}_{(QX)}}{I} (QX)^\dag e^{V_B} (QX) \nonumber \\
&+& c_{(\bar{Q}X)} \frac{{\cal Z}_{(\bar{Q}X)}}{I} 
(\bar{Q}X)^\dag e^{-V_B} (\bar{Q}X) \nonumber \\
&+& c_{(Q\bar{Q})} \frac{{\cal Z}_{(Q\bar{Q})}}{I} (Q\bar{Q})^\dag (Q\bar{Q}), 
\end{eqnarray}
where overall coefficients $c$'s are of order ${\cal O}(1)$ unknown constants. 
The exponential factors for QCD are suppressed. 
$V_B$ is a background vector superfield $U(1)_B$ with a VEV 
$\langle V_B \rangle = \bar{\theta} \sigma^\mu \theta 
\langle A_\mu \rangle, \quad \langle A_\mu \rangle = (\mu/g_B,0,0,0)$. 
$g_B$ is a gauge coupling constant and 
$\mu$ is a chemical potential, which breaks SUSY explicitly. 
Therefore, we assume $\mu \ll \Lambda_{SO(N)}$ so 
that we can make use of exact results of the SUSY gauge theory. 
This VEV of the background vector field provides 
additional tachyonic soft SUSY breaking scalar mass squareds 
as discussed in \cite{HLM}. 
Wave function renomalization constants $Z_i$ are promoted 
to a superfield ${\cal Z}_i$ 
\begin{equation}
{\cal Z}_i = Z_i \left[ 1 - \theta^2 \bar{\theta}^2 m_{{\rm soft}}^2 
\right]
\end{equation}
where $m_{{\rm soft}}$ is a soft SUSY breaking scalar mass in the UV 
and taken to be universal. 
The quantity $I$ is a spurious $U(1)$ symmetry and the RG invariant 
superfield, 
\begin{equation}
\label{invI}
I = \Lambda_h^\dag {\cal Z}^{2T/b_0} \Lambda_h
\end{equation}
where $T$ is the total Dynkin index of the matter fields, 
$b_0$ is the 1-loop beta function coefficient and 
$\Lambda_h = \mu_{UV} {\rm exp}[-8 \pi^2 S(\mu_{UV})/b_0], 
S(\mu_{UV}) = \frac{1}{g^2}\left( 1+ \theta^2 \frac{m_\lambda}{2}\right)
(m_\lambda: {\rm gaugino~mass})$. 
Note that a spurious $U(1)$ transformations are given by 
\begin{eqnarray}
&&Q_r \to e^A Q_r, {\cal Z}_r \to e^{A+A^\dag}{\cal Z}_r, \\
&&S(\mu_{{\rm UV}}) \to S(\mu_{{\rm UV}})-\frac{T}{4\pi^2}A
\end{eqnarray}
where $A$ is a chiral superfield. 

Now, the soft SUSY breaking scalar masses 
for composites with the canonical kinetic term are obtained 
by taking $\theta^2 \bar{\theta}^2$ terms 
in Eq.~(\ref{effkahler}) \cite{footnote1}, 
\begin{eqnarray}
\label{scalar1}
\tilde{m}^2_{(Q^2)} = \tilde{m}^2_{(\bar{Q}^2)} &=& \frac{6N_f + 7}{2(3N_f + 2)} 
m_{{\rm soft}}^2 -\mu^2, \\
\label{scalar2}
\tilde{m}^2_{(X^2)} = \tilde{m}^2_{(Q\bar{Q})} &=& \frac{6N_f + 7}{2(3N_f + 2)} 
m^2_{{\rm soft}} > 0, \\
\label{scalar3}
\tilde{m}^2_{(QX)} = \tilde{m}^2_{(\bar{Q}X)} &=& \frac{6N_f + 7}{2(3N_f + 2)} 
m_{{\rm soft}}^2 - \frac{1}{4} \mu^2. 
\end{eqnarray}
For the case $\epsilon_L = -\epsilon_R$ with vanishing superpotetnial 
in (\ref{superpot}), 
the scalar potetntial consists of only the soft scalar masses 
(\ref{scalar1})--(\ref{scalar3}). 
We then immediately find from (\ref{scalar1}) 
that $\tilde{m}^2_{(Q^2)} = \tilde{m}^2_{(\bar{Q}^2)} < 0$ for 
$\mu > \mu_* (\equiv \sqrt{\frac{6N_f + 7}{2(3N_f + 2)}} m_{{\rm soft}} 
> m_{{\rm soft}})$. 
This means that $\langle Q^2 \rangle, \langle \bar{Q}^2 \rangle \ne 0$ 
in that range of the chemical potential \cite{footnote2}. 

Furthermore, if we take into account the most attractive channel 
hypothesis \cite{MAC}, anti-symmetric part of 
$Q^2 (\afund, \asym, {\bf 1})_{2, 2, 2}, 
\bar{Q}^2 (\fund, {\bf 1}, \overline{\asym})_{-2, 2, 2}$ are 
likely to have VEVs since the force acting on $Q^2$ or $\bar{Q}^2$ 
by one $SO(N)$ gauge boson exchange is attractive. 
On the other hand, the force is replusive in the symmetric case. 
This is our main result that we wish to show. 

We also note that $\mu_* > m_{{\rm soft}}$ implies 
$\mu > m_{{\rm soft}}$ for the condensation of the 
scalar component of the diquark supermultiplet to occur. 
This leads to UV unstable theory 
since the soft SUSY breaking scalar mass squared 
in the UV becomes negative in the presence of the chemical potential, 
$\tilde{m}^2_{Q} = m_{{\rm soft}}^2 - \mu^2 < 0$. 
We therefore add SUSY mass terms 
\begin{equation}
\label{SUSYmass}
W = m_{ij} Q_i \bar{Q}_j, \quad m_{ij} = m \delta_{ij}
\end{equation} 
where the mass is taken to be flavor diagonal 
to preserve the most global symmetry. 
For the UV theory to be stable, 
$\sqrt{m^2 + m_{{\rm soft}}^2} > \mu$ is required. 
The SUSY mass term (\ref{SUSYmass}) can be rewritten in the IR 
as the linear term of composite, 
\begin{equation}
\label{SUSYmass1}
W = m (Q\bar{Q}),
\end{equation}
which breaks SUSY spontaneously. 
We also have to impose the condition $m \ll \Lambda_{SO(N)}$ to 
use the exact results of SUSY gauge theory reliably.

Combining these facts, 
we can obtain 
the diquark supermultiplet 
the scalar component of the diquark supermultiplet condenses 
at some critical value of chemical potenial $\mu_*$. 
We cannot determine exactly where the scalar component of 
the diquark supermultiplet stabilize, 
but in a region where the field VEV is much larger than $\Lambda_{SO(N)}$, 
elementary fields are appropriate variables, 
we know that the potential is stabilized by SUSY mass terms. 
This implies that the VEVs of the scalar component of 
the diquark supermultiplet should 
be stabilized at certain value. 

One may find from (\ref{scalar3}) that composite squarks develop VEVs 
$\langle QX \rangle, \langle \bar{Q}X \rangle \ne 0$ 
for $\mu > 2 \mu_*$ if $2\mu_* < \sqrt{m^2 + m_{{\rm soft}}^2}$. 
However, the scalar component of the diquark supermultiplet 
condenses as far as the chemical potential 
is larger than the critical chemical potential value $\mu_*$. 
In order to analyze the theory in such a case, 
we have to expand our theory around the VEV of the scalar component 
of the diquark supermultiplet condensation. 
Therefore, the above description of the squark condensation 
is left untouched at the present stage. 

\begin{figure}
\begin{center}
\includegraphics[width=\columnwidth,height=2cm]{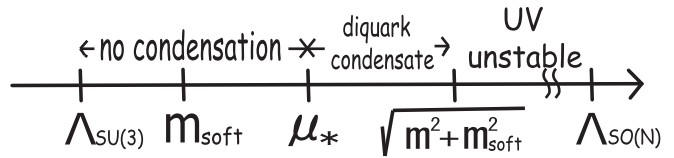}
\end{center}
\caption{\label{fig:scale} Relation among various scales are displayed. 
Horizontal axis means the energy scale. 
The possible range of the chemical potential is 
below $\sqrt{m^2 + m^2_{{\rm soft}}}$. 
There is no condensation when the chemical potential is 
between QCD scale $\Lambda_{{\rm SU(3)}}$ and 
the critical chemical potential $\mu_*$. 
The condensation of the scalar component of the diquark supermultiplet 
occurs when the chemical potential is 
between $\mu_*$ and $\sqrt{m^2 + m^2_{{\rm soft}}}$.
UV theory becomes unstable if the chemical potetntial 
is beyond the scale $\sqrt{m^2 + m^2_{{\rm soft}}}$.} 
\end{figure}

The above argument for the behavior of the scalar component of 
the diquark supermultiplet or 
squark supermultiplet condensation is valid for the vanishing superpotential 
with $\epsilon_L = - \epsilon_R$ in (\ref{superpot}). 
For the case with $\epsilon_L = \epsilon_R$, on the other hand, 
it is found that the scalar potenial is very complicated, 
\begin{eqnarray}
\label{potential}
V &=& \left( \frac{\partial^2 K_{{\rm eff}}}{\partial \Phi_i \partial \Phi^*_j} 
\right)^{-1} \left( \frac{\partial W_{{\rm eff}}}{\partial \Phi_i} \right)
\left( \frac{\partial W^*_{{\rm eff}}}{\partial \Phi^*_j} \right) 
\Lambda^2_{SO(N)} \nonumber \\
&+& \frac{1}{\Lambda^2_{SO(N)}} \left[
\tilde{m}^2_{(Q^2)}|(Q^2)|^2 + \tilde{m}^2_{(\bar{Q}^2)} |(\bar{Q}^2)|^2 
\right. \nonumber \\
&+& \left. \tilde{m}^2_{(X^2)} |(X^2)|^2 
+ \tilde{m}^2_{(QX)} |(QX)|^2 \right. \nonumber \\
&+& \left. \tilde{m}^2_{(\bar{Q}X)} |(\bar{Q}X)|^2 
+ \tilde{m}^2_{(Q\bar{Q})} |(Q\bar{Q})|^2 \right] 
\end{eqnarray}
where $\Phi_i$ denote the scalar component of composite superfields, 
$K_{{\rm eff}}$ is given by (\ref{effkahler}) 
and $W_{{\rm eff}}$ is the sum of the superpotential (\ref{superpot}) 
rewritten in terms of composites and (\ref{SUSYmass1}). 
Therefore, we give here a qualitative discussion 
on the scalar potential behavior 
instead of performing an explicit minimization of the scalar potential. 
Note that F-term contributions to the scalar potential 
in the first line of (\ref{potential}) have a runaway behavior, 
which make the fields VEV away from the origin. 
For $\mu=0$, all SUSY breaking scalar mass squareds are positive, 
which set the fields VEV at the origin. 
Therefore all composites are expected to develop nonvanishing VEVs 
by balancing terms between the runaway potential 
and the SUSY breaking scalar mass terms. 
Even if we take into account that 
the scalar diquark mass squareds become negative for $\mu > \mu_*$, 
qualitative features of phase transition remains unchanged. 
In any case, the case of nonzero superpotential with $\epsilon_L = \epsilon_R$ 
in (\ref{superpot}) is irrelevant to the phase of color superconductivity 
of our interest. 
Even if we compare the vacuum energy in both cases, 
the case with vanishing superpotential seems to be energetically favored.

In summary, motivated by the work of \cite{HLM}, 
we have tried to construct a toy model where gauge noninvariant operators 
are taken into account in SUSY gauge theories. 
We have proposed a SUSY composite model of color superconductivity, 
which is based on an $SO(N)$ gauge theory with $(N-4)$ vector representations 
\cite{IS}. 
Our model is in a confining phase for $SO(N)$ gauge dymanics 
at low energies, 
in which quarks and diquarks are generated dynamically 
as composite fields 
satisfying anomaly matching conditions. 
We have shown that the scalar component of the diquark supermultiplet 
condensate occurs 
when the chemical potential becomes larger than some critical value 
$\mu_{*} = \sqrt{\frac{6N_f+7}{2(3N_f+2)}} m_{{\rm soft}}$. 
Our model is valid 
in the parameter region 
$\Lambda_{{\rm SU(3)}} < \mu < \sqrt{m^2 + m^2_{{\rm soft}}}$, 
where the upper bound is required for the theory to be stable in the UV 
and the lower bound implies that we consider the theory 
where quarks are deconfined. 
Although the model is not fully realistic 
in that the scalar component of diquark supermultiplet
(not diquarks themselves) condense, 
we believe that it well captures some important 
aspects of the diquark condensation behavior and helps our 
understanding for the color superconductivity in real QCD.  
If there is a certain intermediate region of the chemical potential
where quarks are deconfined but not superconducting yet,
owing to the strong quark-quark correlation, the system may be
well described by a compositon of quarks and diquarks. Then
the analysis performed in this paper will help us with
comprehending the behavior of such a system.

As future directions, 
it is interesting to extend our analysis to other flavor cases 
with various phases other than the confining phase. 
In particular, it might be possible to obtain better and more realistic 
understanding for the diquark condensation behavior 
by exploiting Seiberg dual magnetic description \cite{Seiberg}. 
In order to fully understand the phase structure of QCD, 
it is necessary to take into account the finite temperature effects. 
It is therefore indispensable to consider our model extended to 
five dimensional spacetime compactified on $S^1$, 
and then to study the scalar component of the diquark supermultiplet 
condensation behavior 
on the temperature-chemical potential plane as shown in Fig \ref{fig:phase}.

\begin{acknowledgments}
We would like to thank Masashi Hayakawa for valuable discussions 
at various stages of this work. 
We are supported by Special Postdoctoral Researchers Program at RIKEN 
(No.~A12-52040(N.M.) and No.~A12-52010(M.T.)). 
\end{acknowledgments}


\begin{thebibliography}{}
\bibitem{footnote1}
   Strictly speaking, the mass squareds in (\ref{scalar1}) and 
   (\ref{scalar3}) are determinant of mass eigenvalues.
\bibitem{footnote2}
   Throughout this paper, we denote the superfield and 
   its scalar component by using same symbols.
\end{thebibliography}
\end{document}